\documentclass[useAMS,usenatbib]{mn2e}

\usepackage{graphicx}
\usepackage{float}
\usepackage{amssymb,amsmath}
\newcommand{\hstB} {F435W}
\newcommand{\hstV} {F555W}
\newcommand{\hstI} {F814W}
\newcommand{\feh}{\rm{[Fe/H]}}
\newcommand {\Msun}{M$_\odot$}

\title{Intermediate Old Star Clusters in a Young Starburst: The case of NGC~5253}

\author[D. Harbeck et al.]{D. Harbeck$^{1}$, J. Gallagher III$^{2}$,  and D. Crnojevi\'c$^{3}$\\
$^{1}$WIYN Observatory, Tucson, AZ 85719, USA\\
$^{2}$University of Wisconsin, Madison, WI 53706, USA\\
$^{3}$Institute for Astronomy, University of Edinburgh Royal Observatory, Blackford Hill, EH9 3HJ Edinburgh, UK}

\begin{document}

\date{astro-ph version, accepted for publication in MNRAS - The definitive version is available at www.blackwell-synergy.com}

\pagerange{\pageref{firstpage}--\pageref{lastpage}} \pubyear{2002}

\maketitle

\label{firstpage}

\begin{abstract}
We investigate the star cluster population in the outer parts of the starburst galaxy NGC~5253 using archive images taken with the Hubble Space Telescope's Advanced Camera for Surveys. Based on the F415W, F555W, and F814W photometry ages and masses are estimated for bona-fide star cluster candidates. We find three potentially massive ($\ge 10 \time 10^5$ \Msun) star clusters at ages of order of 1-2 Gyr, implying, if confirmed, a high global star formation rate in NGC~5253 during that epoch. This result underlines earlier findings that the current star burst is just one episode in an very active dwarf galaxy.
 \end{abstract}


\section{Introduction}

NGC~5253 is a nearby (D$=$3.5$\pm$0.4~Mpc, McQuinn et al. 2010a) example of a starbursting blue compact dwarf galaxy. Its central regions are dominated by an intense star formation that is embedded in an elliptical main body that resembles a dE galaxy (e.g, van den Bergh 1980; Caldwell \&  Phillips 1989).  The starburst zone has been extensively studied and consists of star clusters located in a gas-rich environment of OB stars and ionized gas (Caldwell \&  Phillips 1989, Beck et al. 1996, Calzetti et al. 1997, Tremonti et al. 2001, Alonso-Herrero et al. 2004). In addition to its unusual morphology, NGC~5253 has an abnormally low metallicity among star bursting galaxies for its stellar mass \citep{dellenbusch2007} and an unusually low HI content (Lopez-Sanchez et al. 2011).

The history of the starburst zone has been studied via a variety of techniques. The star cluster populations suggest an age of $\sim$20~Myr for the central starburst (Harris et al. 2004, Alonso-Herrero et al. 2004, Cresci et al. 2005), a result that needs to be taken in the context of evidence that many of the {central} star clusters may dissolve over short ($\sim$10~Myr) time scales \citep{tremonti2001}.  The star formation history of the galaxy as a whole, however, is critical to understanding the nature of NGC~5253. Van den Bergh (1980) first pointed out the possibility that the peculiar nature of NGC~5253 could result from a close passage near the spiral M~83, its closest giant neighbor in projection. A deep HI study by Lopez-Sanchez et al. (2011) could support this view in revealing the presence of an HI tidal tail that points in the direction of M83. However, as pointed out by Davidge (2007), once computing the 3D distance of NGC~5253 from M~83, a recent past interaction seems to be unlikely, and puts the dwarf in a rather isolated position within the CenA/M83 group. Alternatively, NGC~5253 could have experienced the accretion of a gas-rich companion, as suggested by Kobulnicky \&  Skillman (1995) based on the evidence that its HI gas is rotating about the minor axis of the optical galaxy. The detection of a faint stellar tidal tail around the dwarf starburst NGC~4449 \citep{martinezdelgado2011} emphasizes the potential importance of dwarf-dwarf interaction to trigger a start burst.

Studies of the overall star formation history of NGC~5253 thus hold a potential key to 
its evolutionary history.  Unfortunately the star formation history during the 
past $\sim2$~Gyr has not been determined in any detail. An important step was made by 
McQuinn et al. (2010a,b) who used {\it Hubble Space Telescope} (HST) images to 
analyze the resolved stellar populations of a sample of nearby ($\lesssim 4$~Mpc) star 
bursting dwarfs, including NGC~5253, with the synthetic star formation history recovery 
technique (for a review see, e.g., Cignoni \& Tosi 2010). For color-magnitude diagrams 
(CMDs) that are shallow due to the distance of the studied galaxies, the time 
resolution in recovering star formation episodes is reliable only for the past $\sim 
0.5-1$~Gyr. However, their models indicate the star formation rate (SFR) in NGC~5253, 
aside from the most recent burst, rose steadily during this period. The starburst now 
seen in NGC~5253 is thus not an isolated event, but is part of an extensive star 
forming episode (McQuinn et al. 2010b), although the full duration of this major event 
remains to be established.

Age dating of star clusters has long been recognized as offering a powerful means for exploring the evolutionary histories of galaxies (e.g., Woodley 2010 and references therein). Caldwell \&  Phillips (1989)  mapped star cluster candidates in NGC~5253 using high quality ground-based CCD images obtained with the Blanco 4-m telescope at Cerro Tololo InterAmerican Observatory.  Their catalog contains 121 candidate star clusters, but they were well aware that confusion with stars and galaxies was an issue. Thus while this list contains a number of star clusters, and suggests that clusters are likely to exist in the main stellar body of NGC~5253, it inevitably suffers from substantial contamination.

With the availability of images taken with the HST Advanced Camera for Surveys (ACS), confusion with stars and background galaxies can be greatly reduced. In this paper we { complement earlier work on the central young star cluster population with a photometric study of candidate globular cluster like star clusters in the stellar envelope of NGC~5253}. Although the reality of the star clusters and their abundances must be spectroscopically confirmed, the overall sample contains several relatively certain cluster candidates whose colors are consistent with intermediate ages, { and we will single out the three most massive cluster candidates to infer a potentially high star formation rate in NGC~5253 1-2 Gyr {before the present}; thus this work represents a first step towards a more detailed understanding of star formation history of NGC~5253 extending further than 0.5~Gyr into the past.}

\section{Data, Analysis, \& Cluster Sample}

We base this paper on archive HST observations in two fields of NGC~5253 that were taken by the Advanced Camera for Surveys (ACS) in the F435W, F555W, and F814W passbands. The  observation log is summarized in in Table \ref{tab-obslog}. We will refer to the two pointings as Field I and Field II. The data were originally observed for program 10765, Principal Investigator Andreas Zezas; the date of observation is Dec 27$^{\rm th}$ 2005. We base this perp on the ACS pipeline reduced and drizzle-combined data products from the archive.

\subsection{Star cluster Identification and Photometry}

A Milky Way like Globular Cluster (GC) with a typical half light diameter of order of $3$ pc to $10$ pc will have an apparent size of $\le 0.2$'', to $0.6$'' at the distance of NGC~5253, or $4$ to $12$ ACS pixels. The extent can be larger for the most extended clusters, i.e., massive star clusters at the distance of NGC~5253 will clearly distinguish from point sources (also see Brodie et al 2011). In fact, one would expect to marginally resolve the brightest stars in the most massive  clusters. A cluster with the average luminosity of a Milky Way GC of $M_V=-7.2$\,mag will have an apparent magnitude of $\sim 19.8$ mag and is readily detectable in the available ACS observations. While fuzzy, partially resolved objects are difficult to automatically detect in images, they can be practically identified visually; for this work {we rely solely on the latter detection method since it is the most viable for the limited survey area of the two ACS pointings.}

We create a  candidate input catalog by visually searching bright {, unsaturated,} and resolved objects in $ 31\times31$ pixels median-subtracted F814W-band images. {We inspect the radial profile and FWHM of each potential cluster during this search to distinguish between stars and marginally resolved star clusters; the separation between those would not be obvious form a visual inspection alone.} The median subtraction efficiently removes the diffuse component of the host galaxy and limits the dynamical range in the searched images. We detected a total of 29 candidates in Field I, and additional 2 candidates in the non-overlapping part of Field II. We perform aperture photometry in each of the three passbands on the  drizzled-combined images with the IRAF\footnote{IRAF is distributed by the National Optical Astronomy Observatory, which is operated by the Association of Universities for Research in Astronomy (AURA) under cooperative agreement with the National Science Foundation.} {\tt apphot} package using a 15 pixels aperture radius, and a 10 pixel wide sky aperture. Given the relatively low background din NGC~5253's stellar envelope we chose a single relatively large aperture to measure the flux for all cluster candidates in a consistent manner. The measured flux is transformed into the instrumental HST ST magnitude system  using zeropoints\footnote{Zeropoints for data obtained before July 4, 2006, see ACS Manualhttp://www.stsci.edu/hst/acs/} of z$_{\rm \hstB}$ = 25.168, z$_{\rm\hstV}$=25.688, and z$_{\rm \hstI}$=26.799.

A first attempt to automatically measure the radial structure of the star cluster candidates with the IRAF's radprof tasks turned out to be unreliable (due to the substructure in the star clusters), and we resorted to  manually measuring the full width half maximum (FWHM, i.e., the half-light diameter) of the cluster candidates with IRAF's {\tt imexam} routine, which turned out to be much more robust, while also providing the benefit of visual feedback.

\begin{figure*}
\includegraphics[width=18cm]{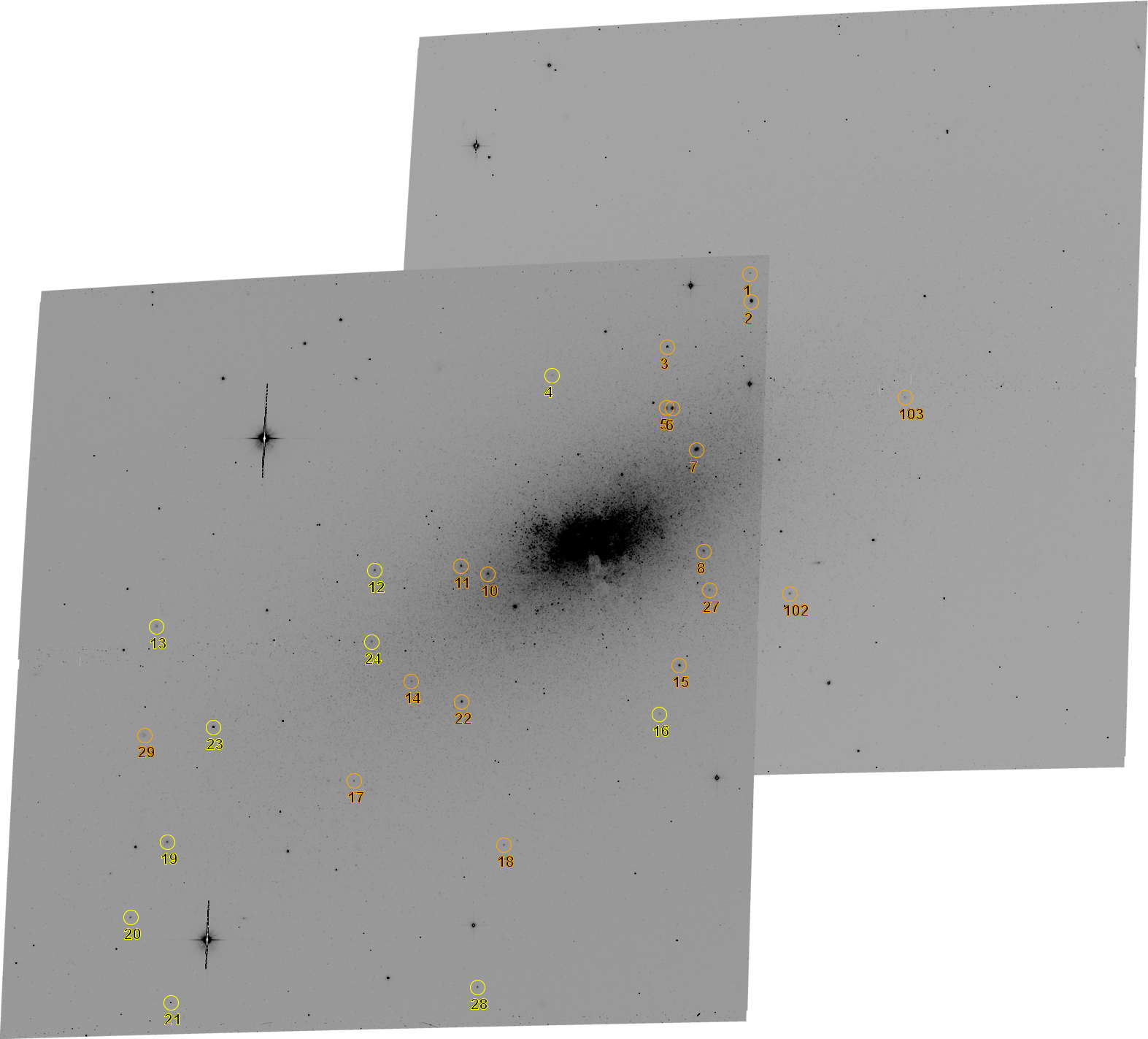}

\caption{Finding chart for star cluster candidates in NGC 5253. The lower-left part of the image is Field I, and the upper right part is Field II. Color-selected cluster candidates are plotted with orange circles, background galaxies are plotted with yellow circles.}
\label{fig-findingchart}
\end{figure*}

\begin{figure}

\includegraphics[width=84mm]{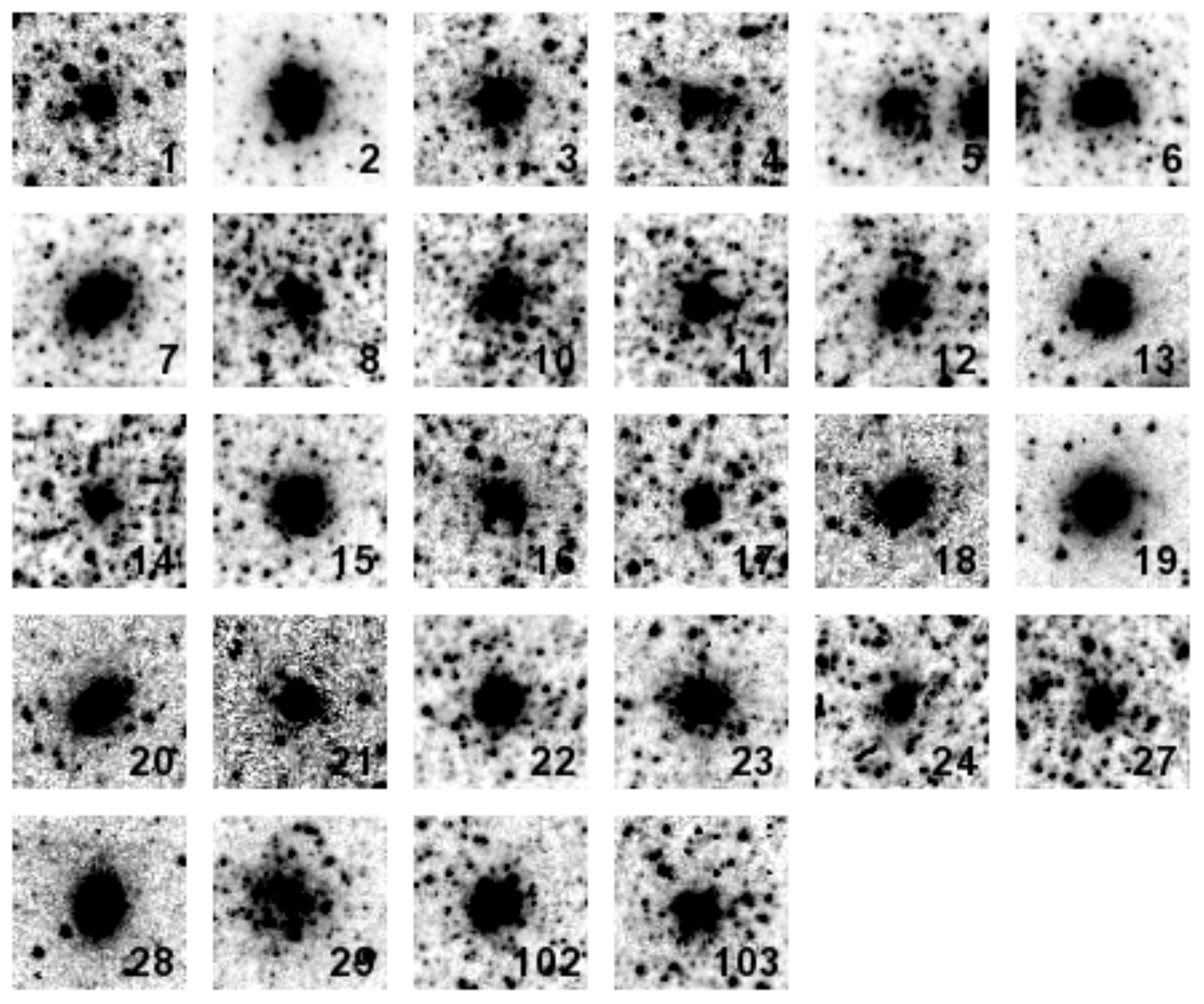}
\caption{Cut-out images for all star cluster candidates. Each image is  64x64 pixels large, or ~2.5" on sky.}
\label{fig-clusterthumbnails}
\end{figure}

\begin{figure*}

\includegraphics[width=18cm]{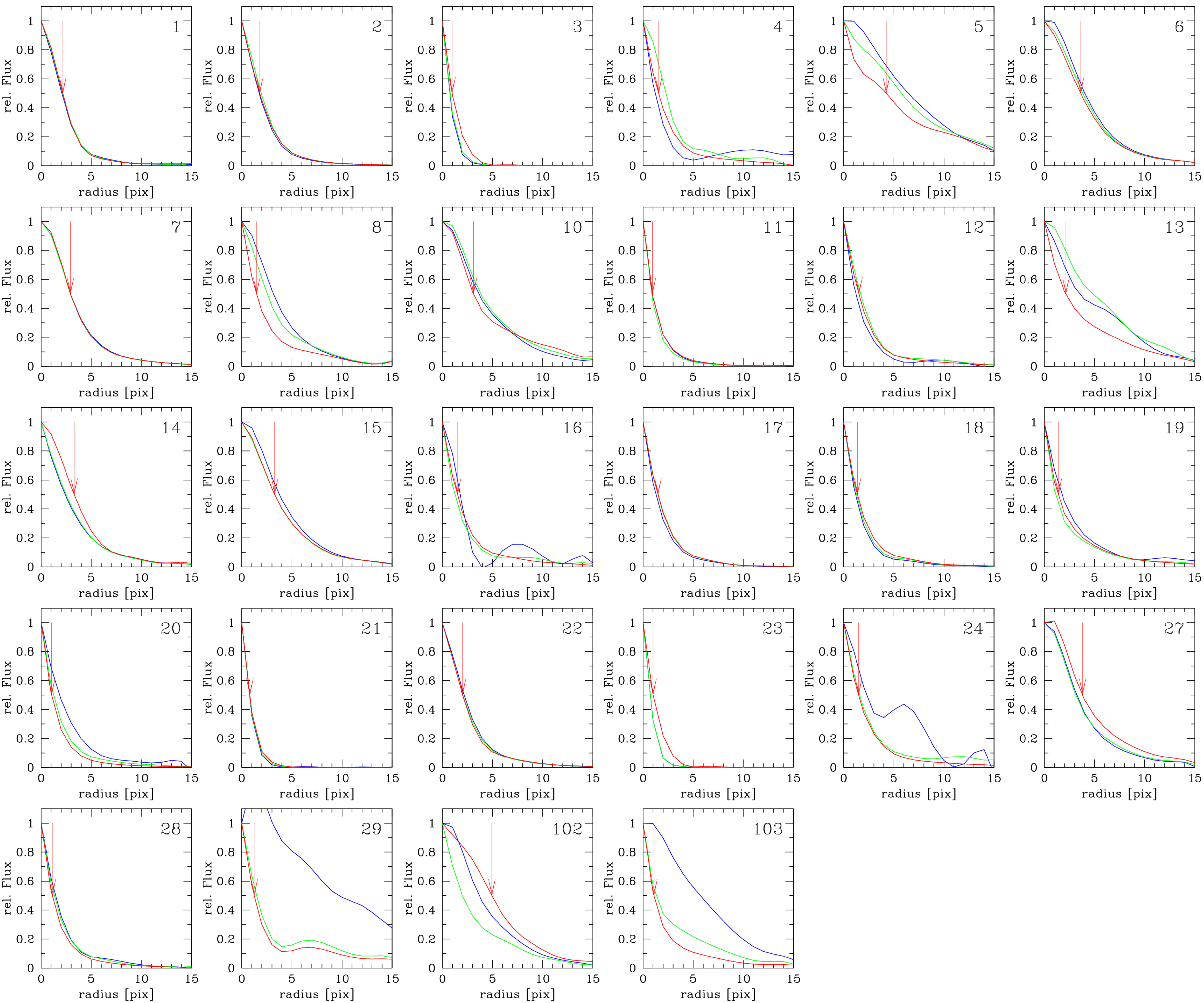}
\caption{Normalized radial profiles of the star cluster candidates in the F435W (blue), F555W (green) , and F814W (red) filters. The arrow marks the half light radius in the F814W band. The classification into clusters and background objects is solely based on a color cut.}
\label{fig-clusterradprof}

\end{figure*}

The photometric and structural properties of { the star cluster candidates} are presented in Table \ref{tab-clusterphotometry}  {, and as no photometric or structural selection has been applied yet, we expect this list to be contaminated by non-cluster objects}. The location of the clusters candidates are marked in the finding chart Fig.~\ref{fig-findingchart}, and cutout images of the star clusters are show in Figure \ref{fig-clusterthumbnails}. We also show the radial profiles in the three colors for all star cluster candidates in Figure \ref{fig-clusterradprof}.

\subsection {Photometry Errors}

A subset of star clusters is located on both observed fields, and we use that sample of objects to gauge the accuracy of the flux measurements. For the bright objects the difference in magnitude between the two observations is of order of $0.01$\,mag, i.e., the error introduced by calibrating and resampling of the drizzled images is marginal.

The second obvious source of photometric errors is subtracting the variable background of the host galaxy. We quantify the systematic error of cluster photometry by comparing different sky subtraction strategies in IRAF's {\tt apphot}. The first approach calculates the {\em mode} in the sky aperture annulus and will  reject bright stars, and thus represent the underlying diffuse galaxy light. The second algorithm calculates the {\em mean} level in the sky aperture and will also include the light of resolved bright stars. As the latter approach will be a better reflection of the underlying galaxy light we will use the {\em mean} background level in the photometry, but note that the background subtraction is subject to stochastic effects in the underlying background stellar population.

We compare the resulting photometry of both background subtraction algorithms in the F814W band, and we find the difference between theses two to be of the order of 0.1 mag; this is the error we assume for the star cluster photometry.

\subsection{Reddening and distance modulus}

We use a global Galactic foreground reddening of $E(B-V) = 0.056$ mag \citep{schlegel1998}, but we note that extinction in the central part of NGC~5253 is highly variable, and an additional error source for the cluster's photometry is differential internal reddening. However, outside of the central region of NGC~5253, internal extinction becomes less important; \citet{calzetti1999} assume a constant extinction of $E(B-V)=0.05$\,mag to account for internal extinction. Since we concentrate on star clusters with significant distances to the galaxy's gaseous center, we  choose to ignore internal extinction, but note that it could still play a role (the dust finger in NGC~5253 should be a warning). In this paper we will not correct the photometry directly, but rather apply extinction corrections to the stellar population models. 

We adopt a distance modulus of ${\rm DM} =27.7$\,mag from the NED data base, which is  consistent with the distance modulus of 27.75\,mag and 27.65\,mag found from the tip of the red giant branch (TRGB) luminosity by \citet{mcquinn2010a} when fitting the CMDs for the high and low surface brightness regions of NGC~5253, respectively.

\subsection{Stochastic effects on cluster photometry and GALEX cross-matching}

Due to the diminishing population of stars at the high mass {,i.e., luminous,} end of the stellar mass function, the {  evolution of a single massive young star onto the red supergiant branch} in a star cluster could { cause the integrated color of that very cluster to falsely appear like that} of an older, more massive star cluster (e.g., \citealt{weidner2004}). According to Fig. 2 of their paper, star cluster with an age older than $10^8$ years are not expected to effected by this confusion.  Near UV photometry of a star cluster is better suited to identify very young star { clusters} than the available ACS photometry {alone}, and we have searched for UV emission by the cluster candidates in an archival Galex near-UV (NUV) image (image stack GI6\_026013\_NGC5253; exposure time 2212.3s; PI: Evan Skillman). We do not find significant near-UV emission in any of the star clusters, but {quantified a detection limit in the UV for the three cluster candidates} \#2, \#6, and \#7. { These three star cluster candidates will be identified in Section 3.2. to be the most massive star clusters in our sample and will be used to infer a star formation rate based on the most massive cluster formed during a star formation event. }

Thus these three candidates deserve increased scrutiny. UV detection in the Galex image is severely limited by a low resolution compared to the ACS images and by the background noise of the underlying galaxy light. We estimate the background noise at 0.005 counts / sec/ pixel and derive a three-sigma detection limited flux  for a point source on top of the background of NGC~5253 at about 0.3 counts / sec, which corresponds to 21.4 mag$_{AB}$. By comparing limits of the the NUV-F435 color with a SPP model, the NUV detection limit constrains the ages of the three massive clusters to be older than $0.5$~Gyr, i.e., for none of the three clusters the optical color should be dominated by a single red super giant star younger than about a hundred Myr.

\section{Properties of Star Clusters}

\begin{figure*}
\begin{minipage}{84mm}
\includegraphics[width=84mm]{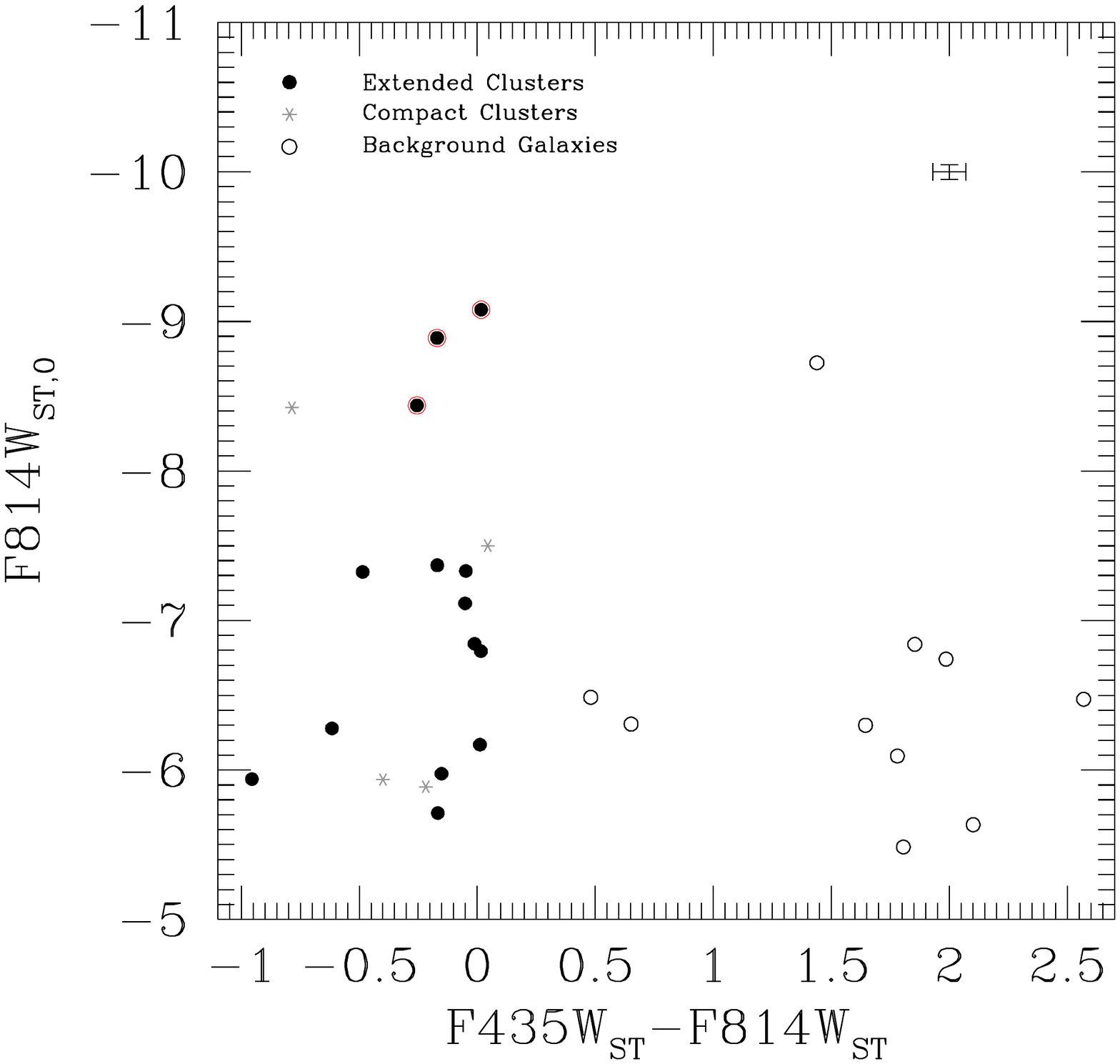}
\end {minipage} 
\begin{minipage}{84mm}
\includegraphics[width=84mm]{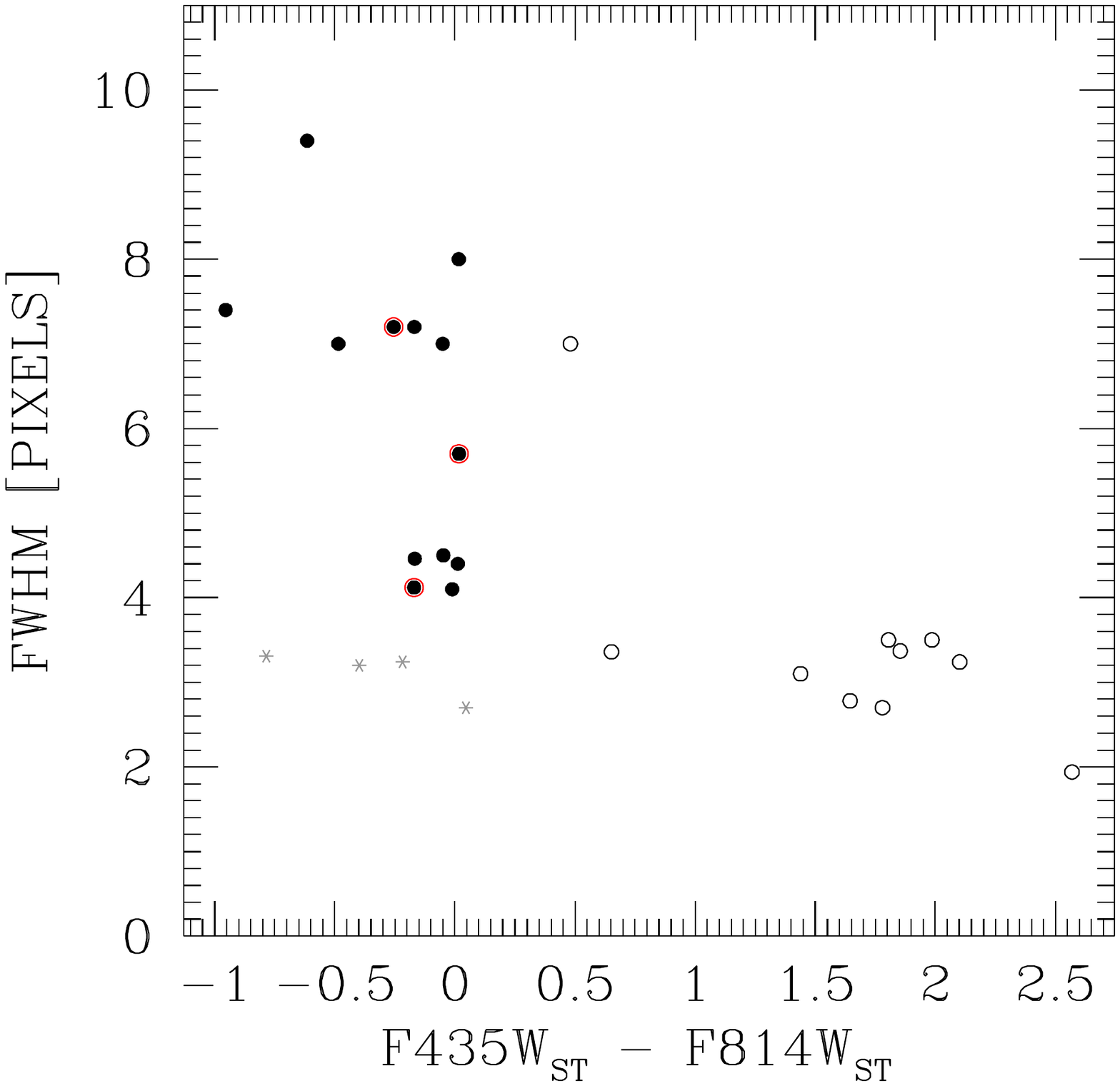}
\end{minipage}
\caption {Left: Color-magnitude diagram of star cluster candidates, where the F814 magnitude is dereddened and corrected for distance. Colors are uncorrected. Right: Full width half maximum {in the F814W band} of cluster candidates shown as a function of their colors. {The most massive }cluster candidates \#2, \#6, and \#7 are marked with a red circle.}
\label{fig-clustercmd}

\end{figure*}

\begin{figure}

\includegraphics[width=84mm]{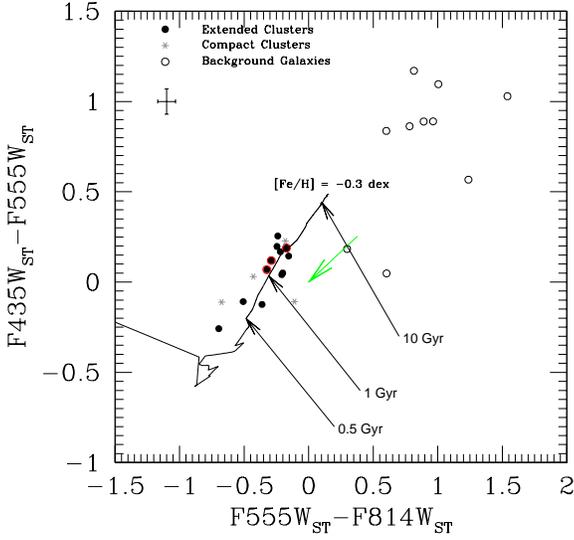}

\caption{Color-Color magnitude diagram of star cluster candidates. A clump of very
red objects (red open circles) is identified as probably background galaxies. A STARBURST99 single stellar population 
evolutionary track for a metallicity of $\feh = -0.3$ dex is plotted on top; the model has been reddened to correct for foreground extinction. The arrow indicates the direction of the reddening correction. Stellar population ages of 0.5 Gyr, 1 Gyr, and 10 Gyr are indicated by the labels. Cluster candidates \#2, \#6, and \#7 are marked with a red circle. }
\label{fig-colcol03}

\end{figure}

In the following we will investigate the age and masses of the star cluster candidates as far as the limitation of three-color photometry allows.  The sample of star cluster candidates is presented in Figure \ref{fig-clustercmd}. We show the CMD of all star cluster candidates in the left plot, and the FWHM of the the candidates plotted versus the cluster's colors on the right side. The colors in these two figures are instrumental ST system magnitudes, whereas in the CMD the F814W magnitude has been dereddened with A$_{\rm I}=0.109$ mag, and a distance modulus of 27.7\,mag has been applied to obtain absolute magnitudes. The cluster candidates separate into two groups in the color space, which we identify as probable star clusters, and red background galaxies. We separate cluster galaxies and likely background objects with a color cut at ${\rm \hstV} - {\rm \hstI} = 0.5$\,mag. This color cut is better justified by the locus of the star cluster candidates when compared to the Starburst99 single-stellar population (SSP) models\footnote{The model was obtained using the GALEV stellar population synthesis model portal \citep{kotulla2009} in the ST mag system} \citep{leitherer1999} in a color-color diagram (see Fig. \ref{fig-colcol03}). We note that a distinction between star clusters and background galaxies  would not be possible based on the structure of the candidates alone (see Fig \ref{fig-clusterradprof}). In Figure \ref{fig-clustercmd}, background objects are plotted with open circles, and compact objects with a half light radius of less than 4 ACS pixels are plotted as asterisks as they might not be star clusters, but other unclear identifications. We choose a FWHM cutoff of 4 pixels for bona-fide star cluster candidates for their expected size range based on the earlier comparison with the Milky Way's GC system and the resolution limit of the ACS images (FWHM of order of 2 pixels). Bona-fide star cluster candidates are plotted with filled circles.

\subsection {Cluster Ages \& Masses}

\begin{figure}

\includegraphics[width=84mm]{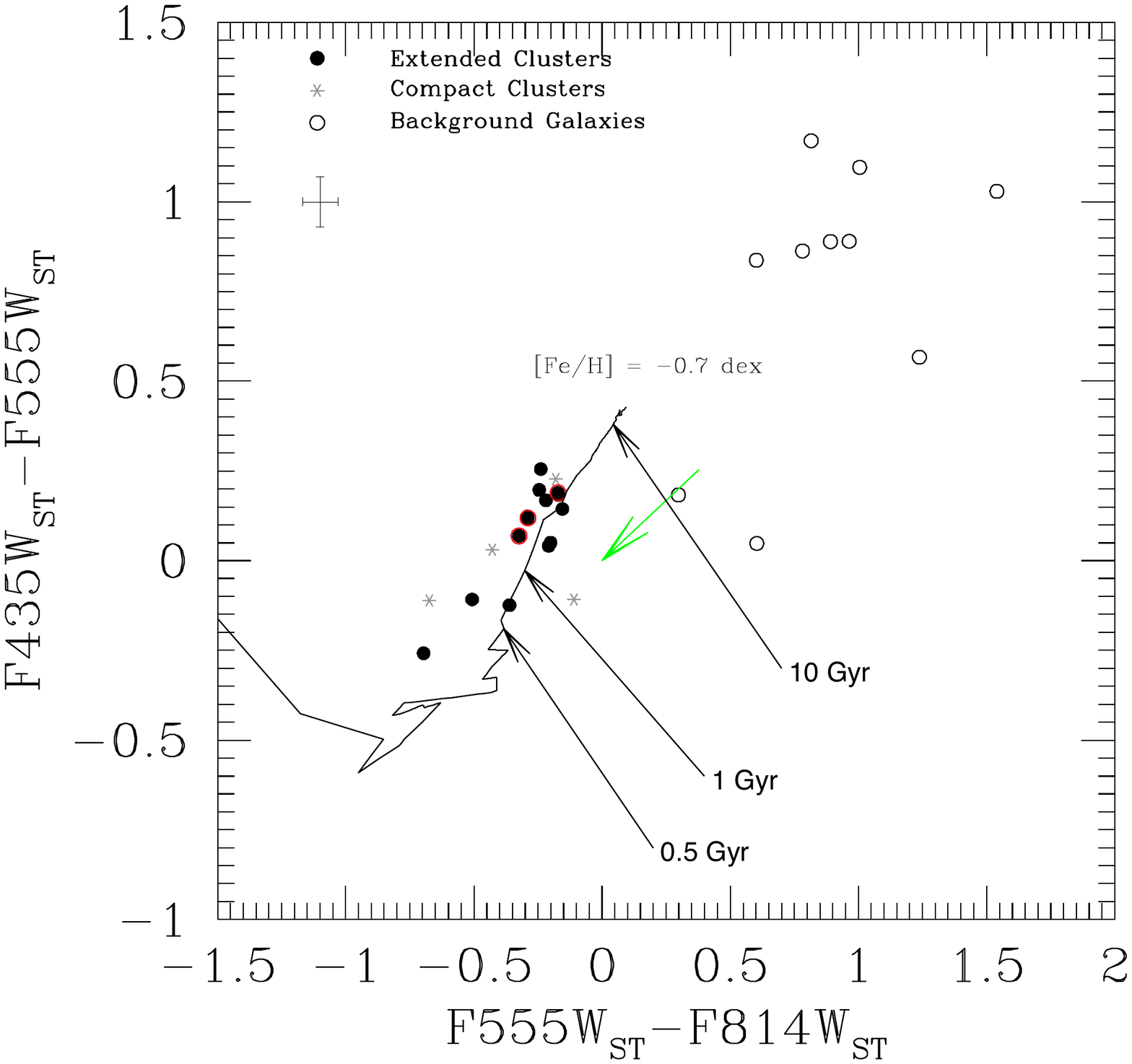}
\caption{Same as Fig. \ref{fig-colcol03}, but with an metallicity of $\feh = -0.7$ dex.}
\label{fig-colcol07}

\end{figure}

\begin{figure}
\includegraphics[width=84mm]{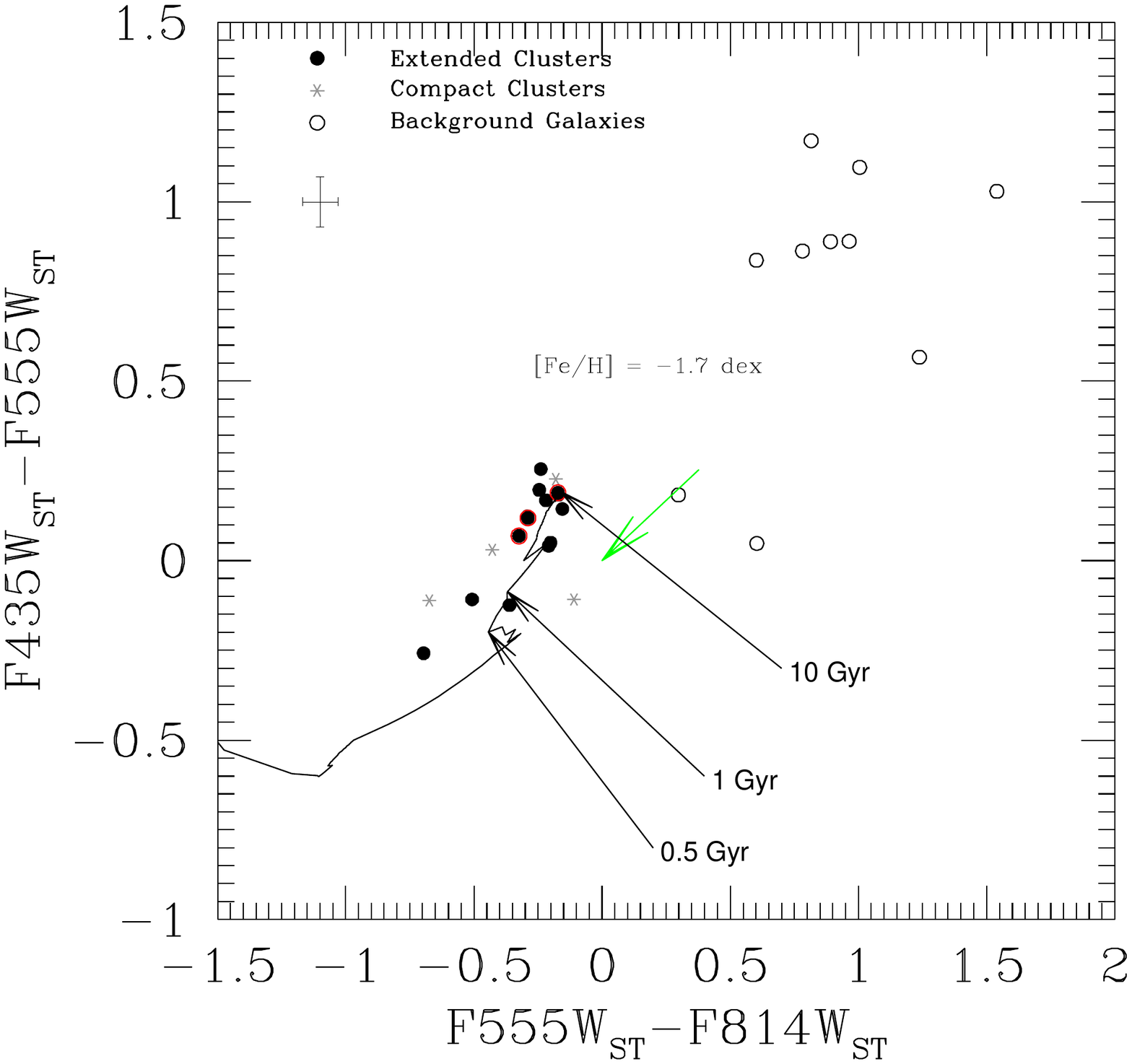}
\caption{Same as Fig. \ref{fig-colcol03}, but with an metallicity of $\feh = -1.7$ dex.}
\label{fig-colcol17}

\end{figure}

In Fig.~\ref{fig-colcol03} through Fig.~\ref{fig-colcol17} we show the instrumental color-color diagrams of the cluster candidates. Each plot is overlayed with a Starburst 99 SSP model with a metallicity of $\feh=-0.3$\,dex, $\feh=-0.7$\,dex, and $\feh=-1.7$\,dex, respectively. The population evolution model has been reddened with $E(B-V) = 0.056$. We approximated  the reddening vector by applying extinction values as defined for the B, V, and I band filters and an G2-star template according to \citet{sirianni2005}. All three SSP models provide an overall viable approximation of the star clusters in the color-color diagrams, underlying the degeneracy in age and metallicity, whereas the two more metal-rich models  {($\feh = -0.3$ dex and $\feh=-0.7$ dex) seem to approximate the entire star clusters candidate population better than the most metal poor model.}

An accurate age determination of the star clusters based on SSP is limited by the degeneracy of age and metallicity in the available colors. Nevertheless, we determine age and mass for each of the star cluster candidates for each of the three metallicities by finding the best fitting model in the sense of the smallest distance between the evolutionary track and the measured three-color photometry. We derive the initial mass of a star cluster by scaling the F814W-band luminosity of a cluster to the best fit SSP model luminosity (the SSP model has been generated for a $10^5$ \Msun initial mass SSP). Formal errors in age are derived by finding the best fit SSP model when both colors are simultaneously varied by $\pm 0.1$~mag according to the photometric uncertainty. 

\begin{figure}

\includegraphics[width=84mm]{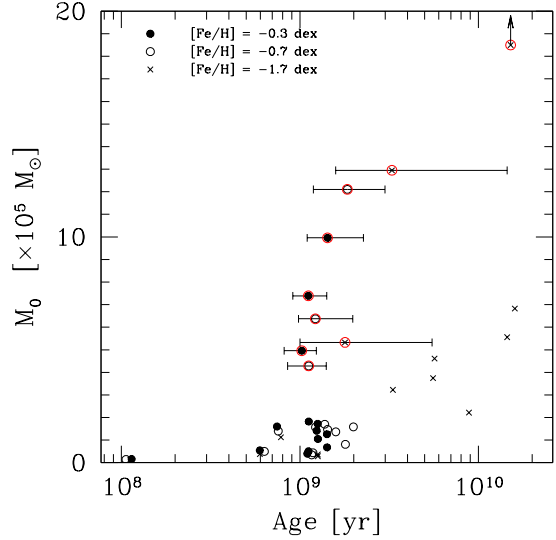}
\caption{Best-fit age and mass estimate  for star cluster candidates with a FWHM of more than 4 pixels.  The ages and masses were estimated
 assuming metallicities of -1.7\,dex (crosses), -0.7\,dex (open circles), and -0.3\,dex (filled circles). The most massive clusters \#2, \#6, and \#7 are marked with a red circle, { and the statistical uncertainty in age (propagated form the photometry, see Sect. 3.1) is shown for these objects. The systematic uncertainty of their ages and masses due to the unknown metallicity is represented by spread of the data points for the three different metallicity models.}}

\label{fig-agemass}

\end{figure}

The resulting masses and ages of bona-fide star clusters are shown in Figure \ref{fig-agemass}; only star cluster candidates with a FWHM $\ge 4$ pixels are shown. We note that the formal uncertainty of an age determination is of order of a a few Gyrs, but the modeling uncertainties are clearly dominating as demonstrated by the different age-mass distribution of the star clusters when different metallicities are assumed.

\subsection {Star formation rate from the maximum mass star clusters. }

The estimated mass and age of the cluster candidates is highly depended on the  assumed metallicity, and the three color photometry does not allow a case by case determination of the metallicity for each cluster. We thus assume a metallicity of $\feh = -0.7$ as the closest approximation among the three models for the entire star cluster population, since the present day gas abundance in NGC~5253 is about 1/4$^{\rm th}$ solar. However, without spectroscopic follow-up observations the remainder of this section remains speculative, especially since the infall of new gas as indicated by \citet{lopez2011} could invalidate any assumed priors such as a strict age-metallicity relation for the stellar populations in NGC~5253.  For the assumed metallicity we find three massive star cluster candidates that are of particular interest in the age range between 1 and 2 Gyrs, and a mass range of $4$ to $12 \times 10^5$ \Msun; these clusters' identifications are \#2, \#6, and \#7. As can be seen from Fig. \ref{fig-agemass}, the position in the age-mass diagram of these clusters remains similar if assuming $\feh = -0.3$.

The maximum mass of a star cluster at a given age interval can be used to infer the star formation rate at that time (Weidner, Kroupa, \& Larsen 2004, Bastian 2008). Using Fig. 3 of their paper, we infer a possible star formation rate of order of 1 to 10 \Msun $/yr$ for the epoch when those star clusters were formed. This compares to a current global star formation rate of 0.2 \Msun$/$yr in NGC~5253 (summarized in \citealt{lopez2011}), and a star formation rate of order of 0.4 \Msun/yr 450\,Gyr ago \citep{mcquinn2010b}, once taking into account the uncertainties of our age determinations.

Regarding the origin of the enhanced star formation epoch about 1 to 2 Gyrs  {in the past}, we can restrict the possible scenarios by computing the physical distances of NGC~5253 to the dominant giants of the CenA/M83 complex. Assuming the TRGB distances computed by Karachentsev et al. (2007), NGC~5253 is located $\sim0.8\pm0.2$~Mpc away from CenA and $\sim1.7\pm0.5$~Mpc away from M83. Even considering the large errorbars, it is rather unlikely that the dwarf has been close to any of the two giants in the past 2~Gyr, while based on the angular separation alone such an interaction would appear feasible. A possibility is that NGC~5253 is simply experiencing a ``gasping'' star formation history, typical of dwarf irregular galaxies (e.g., Marconi et al. 2005), such that the star formation episodes are regulated primarily by its mass rather than by the surrounding environment. We cannot however exclude a possible interaction with other low-mass dwarfs or with the intra-group medium of the CenA/M83 group. On the other hand, the current central starburst could be an independent event triggered by the infall of intergalactic gas, as suggested by \citet{lopez2011}

\section{Summary \& Conclusions}

The availability of ACS F435W, F555W, and F814W imaging of NGC~5253
provides the basis for a new survey to detect star clusters outside of
the starburst core region.  Due, however, to the relatively high stellar
density in the main body of NGC~5253, difficulties still exist in
separating background galaxies and foreground stars from star clusters.
We therefore experimented to determine if quantitative intensity profile
measurements of star clusters, foreground stars, and the best star
cluster candidates could be used to find clusters. Our results for
single bandpass images demonstrate that such differences do not exist at
sufficient high levels to distinguish clusters form contaminating
objects.  On the other hand, star clusters follow a relatively
well-defined locus in the F555W-F814W vs. F435W-F555W STMAG system, so a
combination of structure and color selection has been used to yield a list of likely star cluster candidates.  As shown in Figure~8, the majority of candidate star clusters in the outer parts of NGC~5253 have ages of up to $\sim$2~Gyr for an assumed metallicity of [Fe/H]$= -$0.7. This is chosen to match the abundance of the gas content of NGC~5253. Given the age-metallicity degeneracy, not resolvable with photometric data alone, the derived ages would be older if choosing a lower metallicity value. A spectroscopic study is needed to confirm the clusters and to better constrain their metallicity to fully define their properties. Their presence, however, is consistent with other literature results, showing that NGC~5253 experienced a long term epoch of enhanced star formation within the past $\sim$1~Gyr.

\section*{Acknowledgments}

JSG appreciates partial support for research on starburst galaxies from NSF grant AST0708967 to the University of Wisconsin-Madison. DC acknowledges support from an STFC rolling grant.

Based on observations made with the NASA/ESA Hubble Space Telescope, obtained from the data archive at the Space Telescope Science Institute. STScI is operated by the Association of Universities for Research in Astronomy, Inc. under NASA contract NAS 5-26555.

This research has made use of the NASA/IPAC Extragalactic Database (NED) which is operated by the Jet Propulsion Laboratory, California Institute of Technology, under contract with the National Aeronautics and Space Administration.

\newpage

\clearpage

\clearpage

\clearpage

\clearpage

\begin{table}

\caption{Observations}
\begin{tabular}{lccccl} 
\hline

  Field &
  R.A. (J2000.0) &
  Dec  (J2000.0)&
  Filter &
  T${_{\rm exp}}$ [s] &
  Drizzled data product \\
 
\hline

Field I  &  13 39 59.23  & -31 37 40.7 & F435W & $  880 $ & j9k501020\_drz.fits \\
         &               &             & F555W & $ 1200 $ & j9k501010\_drz.fits \\
         &               &             & F814W & $ 1160 $ & j9k501030\_drz.fits \\
Field II &  13 39 52.24  & -31 39 08.2 & F435W & $  879 $ & j9k501060\_drz.fits \\
         &               &             & F555W & $ 1200 $ & j9k501040\_drz.fits \\
         &               &             & F814W & $ 1200 $ & j9k501050\_drz.fits \\
\hline

\end{tabular}
\label{tab-obslog}

\end{table}

\begin{table}
\centering

\begin{minipage}{140mm}
\caption{Star Cluster Candidates}

\begin{tabular}{lccccccc} 
\hline
{ID} &
 x-center\footnote{Coordinates are in \hstI band image pixels; coordinates are in Field I for IDs $\le 29$, and in Field II for IDs $\ge 100$.} &
 y-center &
 F435W$_{ST}$ [mag]\footnote{The typical systematic error for the photometry is $0.1$ mag (see Section 2.2).} &
 F555W$_{ST}$ [mag]&
 F814W$_{ST}$ [mag]&
FWHM [pixels] \footnote{as measured in the F814W band} &
Cluster / Galaxy\footnote{Classification into star clusters (0) and background galaxies (1) by color separation as discussed in Section 3.}
 \\ \hline

  1 & 4110.06 & 4195.07 & 21.92 & 21.88 & 22.09 &  4.5 & 0  \\ 
  2 & 4117.64 & 4043.19 & 18.74 & 18.62 & 18.91 &  4.1 & 0  \\ 
  3 & 3656.82 & 3792.26 & 18.60 & 18.71 & 19.38 &  3.3 & 0  \\ 
  4 & 3025.6 &  3637 & 24.12 & 23.55 & 22.32 &  3.5 & 1  \\ 
  5 & 3651.61 & 3459.51 & 20.64 & 20.47 & 20.69 &  7.0 & 0  \\ 
  6 & 3685.92 & 3456.6 & 19.11 & 19.04 & 19.36 &  7.2 & 0  \\ 
  7 & 3817.05 & 3231.32 & 18.74 & 18.55 & 18.72 &  5.7 & 0  \\ 
  8 & 3855.92 & 2673.82 & 20.95 & 20.80 & 20.96 &  4.1 & 0  \\ 
 10 & 2674.44 & 2547.97 & 19.99 & 20.12 & 20.48 &  7.0 & 0  \\ 
 11 & 2526.8 & 2592.55 & 20.35 & 20.12 & 20.30 &  2.7 & 0  \\ 
 12 & 2053.68 & 2568.68 & 23.05 & 21.88 & 21.06 &  3.5 & 1  \\ 
 13 & 857.57 & 2262.8 & 21.80 & 21.61 & 21.32 &  7.0 & 1  \\ 
 14 & 2254.81 & 1959.67 & 20.91 & 21.17 & 21.86 &  7.4 & 0  \\ 
 15 & 3722.9 & 2048.28 & 20.26 & 20.14 & 20.43 &  7.2 & 0  \\ 
 16 & 3614.96 & 1782.53 & 24.27 & 23.17 & 22.17 &  3.2 & 1  \\ 
 17 & 1940.46 & 1415.51 & 21.47 & 21.44 & 21.87 &  3.2 & 0  \\ 
 18 & 2761.18 & 1064.39 & 21.70 & 21.80 & 21.91 &  3.2 & 0  \\ 
 19 & 916.859 & 1080.38 & 22.82 & 21.93 & 20.96 &  3.4 & 1  \\ 
 20 & 715.651 & 667.558 & 23.49 & 22.60 & 21.71 &  2.7 & 1  \\ 
 21 & 935.941 & 200.445 & 23.90 & 22.87 & 21.33 &  1.9 & 1  \\ 
 22 & 2530.81 & 1849.43 & 20.42 & 20.23 & 20.47 &  4.5 & 0  \\ 
 23 & 1170.34 & 1710.48 & 20.52 & 19.68 & 19.08 &  3.1 & 1  \\ 
 24 & 2036.42 & 2177.92 & 22.15 & 22.10 & 21.50 &  3.4 & 1  \\ 
 27 & 3888.33 & 2461.27 & 20.91 & 21.02 & 21.52 &  9.4 & 0  \\ 
 28 & 2616.24 & 285.85 & 23.15 & 22.29 & 21.50 &  2.8 & 1  \\ 
 29 & 792.366 & 1664.62 & 21.68 & 21.63 & 21.83 & 20.0 & 0  \\ 
102 & 2258.51 & 1049.08 & 21.02 & 20.77 & 21.01 &  8.0 & 0  \\ 
103 & 2887.87 & 2124.9 & 21.64 & 21.46 & 21.63 &  4.4 & 0  \\

\label {tab-clusterphotometry}
\end{tabular}
\end{minipage}
\end{table}

\label{lastpage}


\begin{thebibliography}{99}

\bibitem[Alonso-Herrero et al.(2004)]{alonso2004} Alonso-Herrero, A., Takagi, T., Baker, A.~J., et al.\ 2004, ApJ, 612, 222

\bibitem[Bastian(2008)]{2008MNRAS.390..759B} Bastian, N.\ 2008, MNRAS, 390, 759 

\bibitem[Beck et al.(1996)]{beck1996} Beck, S.~C., Turner,
J.~L., Ho, P.~T.~P., Lacy, J.~H., \&  Kelly, D.~M.\ 1996, ApJ, 457, 610

\bibitem[Brodie et al.(2011)]{brodie2011} Brodie, J.~P., 
Romanowsky, A.~J., Strader, J., \& Forbes, D.~A.\ 2011, AJ, 142, 199 

\bibitem[Caldwell \&  Phillips(1989)]{caldwell1989} Caldwell, N., \&  Phillips, M.~M.\ 1989, ApJ, 338, 789

\bibitem[Calzetti et al.(1997)]{calzetti1997} Calzetti, D., Meurer, G.~R., Bohlin, R.~C., et al.\ 1997, AJ, 114, 1834

\bibitem[Calzetti et al.(1999)]{calzetti1999} Calzetti, D., Conselice, C.~J., Gallagher, J.~S., III, \& Kinney, A.~L.\ 1999, AJ, 118, 797 


\bibitem[Cignoni \& Tosi (2010)]{cignoni09} Cignoni, M., \& Tosi, M.\ 2010, Advances in Astronomy, 2010 

\bibitem[Cresci et al.(2005)]{cresci2005} Cresci, G., Vanzi, L., \&  Sauvage, M.\ 2005, AaP, 433, 447

\bibitem[Davidge(2007)]{davidge2007} Davidge, T.~J.\ 2007, AJ, 134, 1799


\bibitem[Dellenbusch et al.(2007)]{dellenbusch2007} Dellenbusch, K.~E., Gallagher, J.~S., III, \&  Knezek, P.~M.\ 2007, ApJL, 655, L29

\bibitem[Dong et al.(2008)]{domg2008} Dong, H., Calzetti, D., Regan, M., et al.\ 2008, AJ, 136, 479

\bibitem[Harris et al.(2004)]{harris2004} Harris, J., Calzetti, D., Gallagher,
J.~S., III, Smith, D.~A., \& Conselice, C.~J.\ 2004, ApJ, 603, 503

\bibitem[Karachentsev et al.(2007)]{karachentsev07} Karachentsev, 
I.~D., Tully, R.~B., Dolphin, A., et al.\ 2007, AJ, 133, 504 

\bibitem[Kobulnicky \&  Skillman(1995)]{kobulnicky1995} Kobulnicky, H.~A., \&  Skillman, E.~D.\ 1995, ApJL, 454, L121

\bibitem[Kotulla et al.(2009)]{kotulla2009} Kotulla, R., Fritze, U., Weilbacher,
P., \& Anders, P.\ 2009, MNRAS, 396, 462

\bibitem[Leitherer et al.(1999)]{leitherer1999} Leitherer, C., 
Schaerer, D., Goldader, J.~D., et al.\ 1999, ApJS, 123, 3 

\bibitem[L{\'o}pez-S{\'a}nchez et al.(2011)]{lopez2011}
L{\'o}pez-S{\'a}nchez, {\'A}.~R., Koribalski, B.~S., van Eymeren, J., et al.\ 2011, MNRAS, 1780 {\bf will need to be updated}


\bibitem[Martinez-Delgado et al.(2011)]{martinezdelgado2011} Martinez-Delgado, D., Romanowsky, A.~J., Gabany, R.~J., et al.\ 2011, arXiv:1112.2154 

\bibitem[Marconi et al.(1995)]{marconi95} Marconi, G., Tosi, M., 
Greggio, L., \& Focardi, P.\ 1995, AJ, 109, 173 

\bibitem[McQuinn et al.(2010a)]{mcquinn2010a} McQuinn, K.~B.~W., Skillman, E.~D., Cannon, J.~M., et al.\ 2010, ApJ, 724, 49

\bibitem[McQuinn et al.(2010b)]{mcquinn2010b} McQuinn, K.~B.~W.,
Skillman, E.~D., Cannon, J.~M., et al.\ 2010, ApJ, 721, 297


\bibitem[Schlegel et al.(1998)]{schlegel1998} Schlegel, D.~J., Finkbeiner, D.~P., \& Davis, M.\ 1998,ApJ, 500, 525 

\bibitem[Sirianni et al.(2005)]{sirianni2005} Sirianni, M., et al.\  2005, PASP, 117, 1049 

\bibitem[Thilker et al.(2005)]{thilker2005} Thilker, D.~A., Bianchi, L., Boissier, S., et al.\ 2005, ApJL, 619, L79

\bibitem[Tremonti et al.(2001)]{tremonti2001} Tremonti, C.~A.,
Calzetti, D., Leitherer, C., \&  Heckman, T.~M.\ 2001, ApJ, 555, 322

\bibitem[van den Bergh(1980)]{vandenBergh1980} van den Bergh, S.\ 1980, PASP,
92, 122

\bibitem[Weidner et al.(2004)]{weidner2004} Weidner, C., Kroupa, 
P., \& Larsen, S.~S.\ 2004, MNRAS, 350, 1503 

\bibitem[Woodley(2010)]{woodley10} Woodley, K.\ 2010, Bulletin of 
the American Astronomical Society, 42, 347.05 



\end{thebibliography}
\end{document}